# Vectorial nonlinear optical generation

LI ZHANG,[1, 2] FEI LIN,[2] XIAODONG QIU,[2] AND LIXIANG CHEN[2,*]

[1]*School of Physics and Optoelectronic Engineering, Foshan University, Foshan 528000, China*
[2]*Department of Physics, Collaborative Innovation Center for Optoelectronic Semiconductors and Efficient Devices, and Jiujiang Research Institute, Xiamen University, Xiamen 361005, China*
[*]*chenlx@xmu.edu.cn*

**Abstract:** Nonlinear optical generation has been a well-established way to realize frequency conversion in nonlinear optics, whereas previous studies were just focusing on the scalar light fields. Here we report a concise yet efficient experiment to realize frequency conversion from vector fields to vector fields based on the vectorial nonlinear optical process, e.g., the second-harmonic generation. Our scheme is based on two cascading type-I phase-matching BBO crystals, whose fast axes are configured elaborately to be perpendicular to each other. Without loss of generality, we take the full Poincaré beams as the vectorial light fields in our experiment, and visualize the structured features of vectorial second-harmonic fields by using Stokes polarimetry. The interesting doubling effect of polarization topological index, i.e., a low-order full Poincaré beam is converted to a high-order one are demonstrated. However, polarization singularities of both C-points and L-lines are found to keep invariant during the SHG process. Our scheme can be straightforwardly generalized to other nonlinear optical effects. Our scheme can offer a deeper understanding on the interaction of vectorial light with media and may find important applications in optical imaging, optical communication and quantum information science.

## 1. Introduction

Second harmonic generation (SHG), as the first nonlinear optical phenomenon observed in the experiment, can be traced back to the seminal work about frequency converting from 694.3nm to 347.2nm in crystalline quartz in 1961 [1]. In the following year, Kleinman presented the electromagnetic theory of nonlinear dielectric polarization to explain SHG phenomenon [2]. In 1968, Bloembergen *et al.* discovered the SHG effect on the semiconductor-air and metal-air interfaces [3]. As polarization-dependent SHG signals could provide useful information of the structural properties, SHG has become a standard spectroscopic tool to characterize materials [4,5]. Later, inspired by this attractive feature, Freund developed a kind of biological SHG imaging, i.e., SHG microscopy, to investigate the polarity of collagen fibers in rat tail tendon [6]. Since then, owing to no energy deposition to its interacted matters and localized excitation, SHG microscopy has been utilized for clinical imaging purpose to substantially reduce photobleaching and phototoxicity relative to fluorescence methods [7,8]. Nowadays, it has become a mature imaging technology and is widely used in biology and medicine fields [9]. Except for these applications in spectroscopy, SHG process was also wildly used for new-wavelength laser source [10], manipulation and generation the optical vortices [11], and optical image processing [12,13]. And recently, SHG effect was extended to cover the X-ray [14]. We note that all of the above works are almost focusing on the interaction of scalar light fields with nonlinear media.

Recently, vector beams, possessing the spatially inhomogeneous polarization states in the cross section of light, have attracted tremendous attention in optical communication [15,16], optical micromanipulation [17] and high-numerical-aperture focusing [18]. Their polarization features have been widely explored, e.g., Dennis *et al.* investigated the polarization singularities [19,20], and Milione *et al.* focused on the Pancharatnam-Berry phase of high order vector beams [21,22]. Some researchers have also investigated the SHG process of the vector beams. In 2006, Carrasco *et al.* elaborated theoretically the SHG for vector Gaussian beam [23], then, the SHG process of vortex beams with radial or azimuthal polarization has also been explored [24,25]. Later, SHG about cylindrical vector beam was extended to frequency-tunable process [26]. Besides, Freund investigated the influence of the singularities contribution of vector beams on the generated SHG light field [27], and Stanislovaitis *et al.* studied carefully the conservation of topological charge during the SHG process [28]. Recently, we also visualized the hidden topological structures of a full Poincaré (FP) beam via type-II SHG in the KTP crystals [29].

However, we note that, due to the type-II phase matching condition, the vectorial feature of the fundamental light with spatially inhomogeneous polarization cannot be well maintained after SHG process, namely, the generated SHG beam becomes merely a scalar light field with a uniform polarization state. We also note that the polarization state of light usually plays an irreplaceable role in optical manipulation [30], optical communication [31], and lithography [32]. In this regard, how to maintain the vectorial nature of light fields during the process of SHG is meaningful and desirable from both theoretical and applied points of view. Here, by adopting two cascading type-I phase-matching BBO crystals, whose fast axes are just configured to be perpendicular to each other, we present an experimental scheme to realize SHG from infrared vector fields to visible vector fields. Without loss of generality, we choose the FP beams [33-36], possessing spatially varying polarization states covering the entire Poincaré sphere, to show the feasibility of our scheme. As the horizontal and vertical polarization components of fundamental FP beams participate the SHG process in two cascading BBO crystals independently, the generated light field, as a coherent superposition of the vertical and horizontal SHG light waves, still possess the vectorial feature, i.e., spatially inhomogeneous polarization states. The interesting polarization structures and singularities are further revealed and visualized by measuring the Stokes parameters.

## 2. Theory

The arbitrary-order FP beams can be constructed by using a superposition of a fundamental Gaussian beam and an arbitrary-order LG beams bearing orthogonal polarizations, mathematically expressed as [29,33]:

$$\vec{E}(r,\varphi) = A \cdot \mathrm{LG}_0^0(r,\varphi)\hat{e}_1 + B \cdot \mathrm{LG}_0^l(r,\varphi)\hat{e}_2, \qquad (1)$$

in the cylinder coordinates, where $\hat{e}_1$ and $\hat{e}_2$ represent the unit vectors of left- and right-hand circular components, respectively, $A$ and $B$ are two controllable parameters used to regulate the polarization distribution of the FP beam. And the $l$-th LG beam $\mathrm{LG}_0^l(r,\varphi)$ can be described as,

$$\mathrm{LG}_0^l(r,\varphi) = E_0 r^l L_0^l(\frac{2r}{w^2})\exp(-\frac{r^2}{w^2})\exp(il\varphi), \qquad (2)$$

Where $\varphi$ is the azimuthal angle, $w$ is the beam waist, $L_0^l(\cdot)$ is the associated Laguerre polynomials, and $l$ is an integer. In [29], we presented a stable yet flexibly controlled method to generate FP beams by the combination of the half-wave plate (HWP), quarter-wave plate (QWP) with its fast axis orienting at 45°, and the vortex half wave plate (VHWP). The obtained light fields can be described as,

$$\mathrm{LG}_0^0(r,\varphi)\begin{pmatrix}1\\0\end{pmatrix} \to \frac{(1+i)\cos(2\alpha)}{2}\mathrm{LG}_0^0(r,\varphi)\begin{pmatrix}1\\i\end{pmatrix} + \frac{(1-i)\sin(2\alpha)}{2}\mathrm{LG}_0^2(r,\varphi)\begin{pmatrix}1\\-i\end{pmatrix}, \quad (3)$$

where $\alpha$ denotes the orientation of fast axis of half-wave plate (HWP), which can control the polarization distribution of FP beams. Here, we also adopt this method to generate the fundamental FP beams.

As is well known, due to the phase matching condition, SHG merely yields the scalar beams, i.e. light fields with uniform polarization distribution. From Eq. (1), one can see that the FP beams are the superposition of two orthogonal polarization components, which inspires us to construct the SHG process for each polarization component independently. Thus, we use the QWP with its fast axis orienting at $45°$ to decompose the FP beam into the horizontal and the vertical components:

$$\mathrm{LG}_0^0(r,\varphi)\begin{pmatrix}1\\0\end{pmatrix} \to i\cos(2\alpha)\mathrm{LG}_0^0(r,\varphi)\begin{pmatrix}1\\0\end{pmatrix} - i\sin(2\alpha)\mathrm{LG}_0^2(r,\varphi)\begin{pmatrix}0\\1\end{pmatrix}, \quad (4)$$

We direct these two polarization components pass through two cascading BBO crystals, whose fast axes are just configured to be perpendicular to each other. Besides, both crystals are cut for type-I phase matching to implement the SHG process. Thus we know that in the first BBO crystal only the horizontal polarization component in Eq. (4) participates the process of SHG, generating a frequency-doubled light field of vertical polarization. While in the second BBO crystal, only the vertical polarization components in Eq. (4) participates SHG, yielding a frequency-doubled light field of horizontal polarization. Under the paraxial approximation and the phase-matching condition, the process of frequency doubling in these two crystals can be described, respectively, by the following wave coupling equations,

$$\frac{dE_{SH}^V}{dz} = \frac{i\omega_{SH}^2 d_{eff}}{k_{SH} c^2} E_H^2, \quad (5a)$$

$$\frac{dE_{SH}^H}{dz} = \frac{i\omega_{SH}^2 d_{eff}}{k_{SH} c^2} E_V^2, \quad (5b)$$

where $d_{eff}$ is the effective nonlinear coefficient, $\omega_{SH}$ and $k_{SH}$ are the angular frequency and wave vector of the SHG beam, $c$ is the velocity of light, $E_H$ and $E_V$ represent the horizontal and vertical polarization fundamental light fields, respectively. And under the small-signal approximation, the SHG light fields of orthogonal polarizations can be approximately expressed as $E_{SH}^V \propto E_H^2$ and $E_{SH}^H \propto E_V^2$, respectively. After SHG, another complementary QWP with its fast axis orienting at $45°$ is arranged to recombine $E_{SH}^V$ and $E_{SH}^H$. Accordingly, we have the output SHG beams,

$$\mathrm{LG}_0^0(r,\varphi)\begin{pmatrix}1\\0\end{pmatrix} \to -\cos^2(2\alpha)\left[\mathrm{LG}_0^0(r,\varphi)\right]^2\begin{pmatrix}1\\i\end{pmatrix} + \sin^2(2\alpha)\left[\mathrm{LG}_0^2(r,\varphi)\right]^2\begin{pmatrix}1\\-i\end{pmatrix}, \quad (6)$$

which clearly reveals the acquisition of a higher topological index, in relative to that of the fundamental wave described by Eq. (3). In other words, our scheme can easily realize the frequency conversion from one FP beam to another high-order FP beam, i.e., the vectorial feature can be well preserved during the process of SHG.

## 3. Experimental setup

We display our experimental setup in the Fig.1. After collimated and expanded by a telescope, the horizontal polarized 1064 *nm* laser beams with power 200 *mW* successively pass through a half-wave plate (HWP$_1$,1064 *nm*), a quarter-wave plate (QWP$_1$, 1064 *nm*) with its fast axis orienting at 45°, a vortex phase plate with unit topological charge (VPP, RPC Photonics), and a vortex half wave plate (VHWP, Thorlabs, WPV10L-1064) to produce the desired fundamental FP beams by adjusting the fast axis orientation of HWP$_1$ [29], see inset (a) of Fig. 1 for example. Then another quarter-wave plate (QWP$_2$, 1064 *nm*) with its fast axis orienting at 45° is used to decompose the generated FP beams into two orthogonal components, e.g., the horizontal and vertical polarization, respectively. These two components are imaged via a 4*f* imaging system ($f_1 = 750mm, f_2 = 500mm$) onto two cascading BBO crystals ($8 \times 8 \times 2mm$), whose fast axes are set perpendicular to each other. It was noted that we use two thin crystals such that the walk-off effect could be eliminated effectively. As a result, we implement the SHG process described by Eqs. (5a) and 5(b). After a third quarter-wave plate (QWP$_3$, 532 *nm*), the generated SHG beams of horizontal and vertical polarizations superpose to form the desired FP beams finally, as illustrated in the inset (b) of Fig. 1.

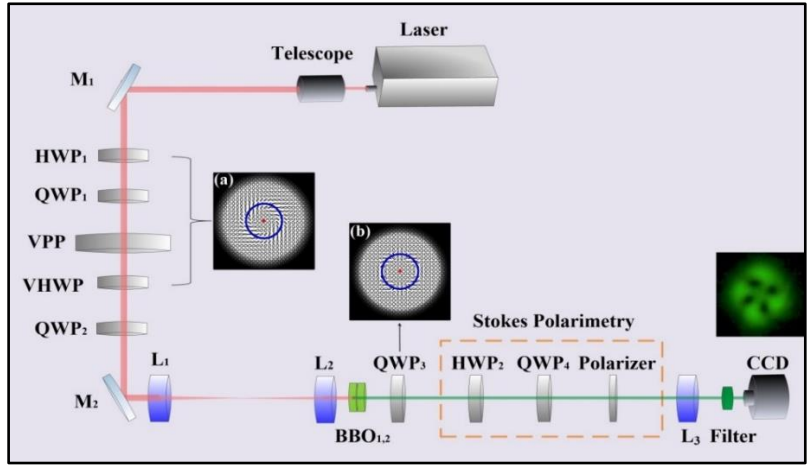

Fig. 1. Experimental setup for SHG of FP beams in BBO crystals.

In order to exactly characterize the polarization states of SHG beams, we conduct the standard Stokes polarimetry with the following HWP$_2$, QWP$_4$ and Polarizer, as shown in Fig. 1. Theoretically, the Stokes parameters can be expressed as [37-39]

$$\begin{cases} s_0 = I_H + I_V \\ s_1 = I_H - I_V \\ s_2 = I_D - I_A \\ s_3 = I_R - I_L \end{cases} \quad (7)$$

where $I_{(\bullet)}$ represents the light intensity of horizontal (H), vertical (V), 45° (D), 135° (A), right-hand (R) or the left-hand (L) polarization component of the SHG beam. Accordingly, one can derive the polarization distribution of an optical field in experiment via measuring the intensity of the aforementioned six components. For this, we arrange the fast axes of HWP$_2$, QWP$_4$ and

Polarizer as shown in the Table 1, where the symbol "-" denotes the unnecessary use of the optical elements. With a CCD camera (Thorlabs, DCU224C), we can easily obtain the spatially varying polarization states point-by-point across the optical field.

Table 1. The settings of the optical elements in the setup to detect the states of polarization

| States<br>Elements | H | V | A | D | R | L |
|---|---|---|---|---|---|---|
| HWP$_2$ | - | - | 22.5° | 22.5° | - | - |
| QWP$_4$ | - | - | - | - | 45° | 45° |
| Polarizer | 0° | 90° | 0° | 90° | 0° | 90° |

## 4. Experimental results and analysis

In our first set of experiment, we verify the effectiveness of the Stokes polarimetry by preparing the fundamental FP beams in the cases of $\alpha = 15°$ and $30°$, see Eqs. (3) and (4). After the cascading SHG process and the Stokes polarimetry, the experimental observations of the six aforementioned polarization components are presented in the first rows of both Fig. 2 and Fig. 3. We can see that the four linear polarization components, H, V, A and D, all exhibit the interference patterns between the fundamental mode and high-order mode, thus, generating four symmetrical off-axis optical vortices all with a single topological charge. While the two circular polarization components are purely fundamental mode for L and high-order mode for R, which conforms to the theoretical simulations shown by the second rows of both Fig. 2 and Fig. 3. According to Eq. (7), we can calculate straightforwardly the Stokes parameters, as shown in the third row. For comparison, we also present the theoretical simulations, as shown by the fourth rows of both Fig. 2 and Fig. 3, from which the good agreement can be seen clearly.

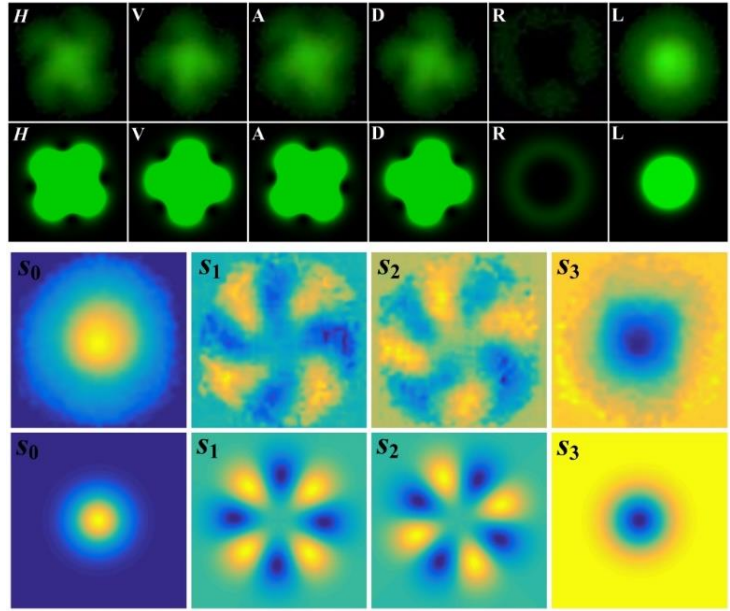

Fig. 2. Polarization and Stokes components for an output SHG light field with $\alpha = 15°$. The first and second rows show the experimental and theoretical results for the six polarization components respectively. The third and fourth rows show the experimental and theoretical Stokes components, respectively.

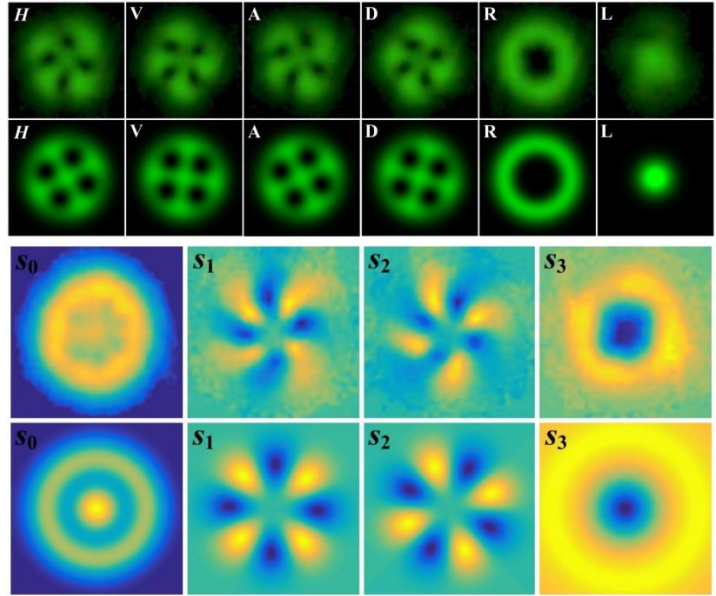

Fig. 3. Polarization and Stokes components for an output SHG light field with $\alpha = 30°$. The first and second rows show the experimental and theoretical results for the six polarization components respectively. The third and fourth rows show the experimental and theoretical Stokes components respectively.

To have a more institutive picture of the spatially varying polarization states, we also plot the polarization ellipse from point to point, by means of the rectifying phase $\chi$ as $\chi = \arcsin(s_3/s_0)/2$ and the orientation angle $\psi$ of local polarization ellipses as $\psi = \arg(s_1 + is_2)/2$ [19,20]. According to Eq. (7) and the experimental results of Figs. 2 and 3, we calculate and visualize the polarization distributions of the SHG beams, as are shown by the bottom panel of Fig. 4. For comparison, we present the simulation results of polarization distribution for both the fundamental and SHG light waves in the top and middle panels, respectively. For the two particular cases: $\alpha = 0°$ and $\alpha = 45°$, since the fundamental beams degrade into the purely circularly polarized LG beam, the generated SHG beams are merely scalar. Of special interests are the rest cases. One can see that, for both the fundamental and SHG light fields, their polarization states across the transverse plane can cover the entire surface of the Poincare sphere. In other words, the vectorial feature of the FP beams can be well maintained in our scheme during the process of SHG. Besides, it is worth noting that the polarization distribution of fundamental beams has the "spiral" structure possessing the topological index $T=2$, which is just equal to the topological charge of the high-order LG mode in Eq. (3). While for the generated SHG light field, the polarization distribution of SHG beams exhibits a Daisy mode and has a topological polarization index of $T=4$, which is twice of that of the fundamental beams. That is to say, SHG doubles not only the frequency of light wave but also the topological polarization index, i.e., a second-order infrared FP beam can be converted to a fourth-order visible one. By comparing the middle and bottom panels, one can see that the experimental results are in good agreement with the numerical simulations.

Furthermore, to describe the polarization singularities [40-43], we recall the rectifying phase $\chi$ and the orientation angle . Namely, the points with $\psi = 0$ denote where the polarization direction is undefined, called C-points. Meanwhile, the points with $\chi = 0$ denote where the polarization handedness is undefined, called L-lines. Here, the C-points marked with red points and the L-lines marked by white circles are plotted in Fig. 4. Both fundamental and SHG FP beams have the same C-point singularity centered at the beam axis, which remains stable as changes. Besides, even though the polarization distribution of SHG beams is different from the fundamental beams with "spiral" pattern, the L-lines behave in a similar way to that of the fundamental wave, i.e., the L-circles shrink as is changing from 0° to 45°, while expand as from 45° to 90°. It should be emphasized that L-lines in the experimental cases of 15° and 75° are located in the margin of patterns, where the intensity is quite weak, so they cannot be plotted. Thus another feature of the LP beam's SHG in our scheme is that regardless of the doubling the topological charge, the polarization singularities, i.e., the C-points and L-lines both keep invariant during the process of SHG.

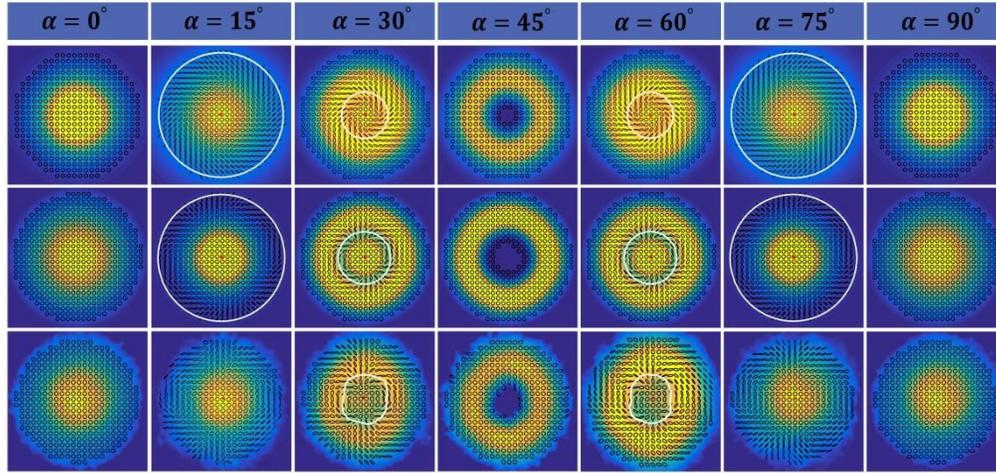

Fig. 4. Polarization distribution of SHG beams with 532nm at $\alpha = 0°$, 15°, 30°, 45°, 60°, 75°, 90°. Top panel: numerical simulations of polarization distribution of fundamental beams; Middle panel: experimental results of polarization distribution of SHG beams; Bottom panel: numerical simulations of polarization distribution of SHG beams.

Within any transverse plane of a paraxial optical field it is useful to characterize the smoothly varying polarization by streamlines oriented along the major axis of the polarization ellipse [40,41]. For an easy visualization, we also plot in Fig. 5 the streamlines around each C-point that is mathematically corresponding to the theoretical and experimental results at $\alpha = 30°$, e.g., the third column of Fig. 4. For the fundamental FP beam of Fig. 5(a), the "spiral" structure of polarization states can be seen intuitively. While for the SHG FP beams, we can see that the spatially varying polarization states exhibit an interesting structure, appearing like the distribution of electric field emitted by an electric dipole consisting of two point electric charges with opposite polarity. Besides, the C point in Fig. 5(b) or 5(c) can be classified by the streamlines of its immediate surroundings into two "lemons", e.g., appearing like the north and south poles, respectively.

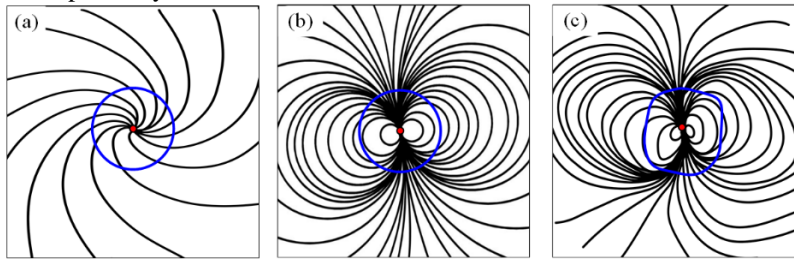

Fig. 5. The streamlines around C-points at $\alpha = 30°$. (a) Numerical result for fundamental FP beam; (b) and (c) Numerical and experimental results of SHG FP beam. Blue lines and red points denote the L-lines and C-points, respectively.

## 5. Conclusion

In summary, we have realized the vectorial second-harmonic generation from infrared FP beams to visible FP beams, based on two cascading type-I phase matching BBO crystals with their fast axes are configured perpendicular to each other. Our experimental observation reveals the interesting doubling effect of topological index, i.e., a low-order FP beam can be effectively converted to a high-order one. Furthermore, the polarization singularities of FP beams, such as C-points and L-lines, both can keep invariant during SHG. This work may offer a deeper investigation on the interaction of vectorial light fields with media and find potential applications in optical imaging and material characterization.

## Funding

This work at Xiamen University is supported by the National Natural Science Foundation of China (91636109, 11474238), the Fundamental Research Funds for the Central Universities at Xiamen University (20720160040), the Natural Science Foundation of Fujian Province of China for Distinguished Young Scientists (2015J06002), and the program for New Century Excellent Talents in University of China (NCET-13-0495). The work at Foshan University is supported by the National Natural Science Foundation of China (11604050, 61575041), Science and Technology Planning Project of Guangdong Province (2016B010113004) and Natural Science Foundation of Guangdong Province (Grant No. 2018A030313347).